# High open-circuit voltage in transition metal dichalcogenide solar cells

Simon A. Svatek[1], Carlos Bueno-Blanco[1], Der-Yuh Lin[2], James Kerfoot[3], Carlos Macías[1], Marius H. Zehender[1], Ignacio Tobías[1], Pablo García-Linares[1], Takashi Taniguchi[4], Kenji Watanabe[5], Peter Beton[3], Elisa Antolín[1]*

1. Instituto de Energía Solar, Universidad Politécnica de Madrid, Avenida Complutense 30, 28040 Madrid, Spain

2. Department of Electronics Engineering, National Changhua University of Education, Changhua 50007, Taiwan

3. School of Physics and Astronomy, University of Nottingham, Nottingham NG7 2RD, U.K.

4. International Center for Materials Nanoarchitectonics, National Institute for Materials Science, 1-1 Namiki, Tsukuba 305-0044, Japan

5. Research Center for Functional Materials, National Institute for Materials Science, 1-1 Namiki, Tsukuba 305-0044, Japan

*e-mail: elisa.antolin@upm.es

**The conversion efficiency of ultra-thin solar cells based on layered materials has been limited by their open-circuit voltage, which is typically pinned to a value under 0.6 V. Here we report an open-circuit voltage of 1.02 V in a 120 nm-thick vertically stacked homojunction fabricated with substitutionally doped $MoS_2$. This high open-circuit voltage is consistent with the band alignment in the $MoS_2$ homojunction, which is more favourable than in widely-used TMDC heterostructures. It is also attributed to the high performance of the substitutionally doped $MoS_2$, in particular the p-type material doped with Nb, which is demonstrated by the observation of electroluminescence from tunnelling graphene/BN/$MoS_2$ structures in spite of the indirect nature of bulk $MoS_2$. We find that illuminating the TMDC/metal contacts decreases the measured open-circuit voltage in $MoS_2$ van der Waals homojunctions because they are photoactive, which points to the need of developing low-resistance, ohmic contacts to doped $MoS_2$ in order to achieve high efficiency in practical devices. The high open-circuit voltage demonstrated here confirms the potential of layered transition-metal dichalcogenides for the development of highly efficient, ultra-thin solar cells.**

Transition metal dichalcogenides (TMDCs) have recently attracted great attention for their potential application in ultra-thin opto-electronic devices because their layered nature provides tuneable, thickness-dependent band gap energies[1] and self-passivated surfaces.[2] Photovoltaic devices based on van der Waals junctions made of TMDCs have demonstrated remarkably high photocurrents for ultra-low thicknesses,[3,4] holding out the promise of TMDCs being chemically stable, non-toxic absorber materials that will enable highly efficient ultra-thin solar cells. However, the open-circuit voltages ($V_{OC}$) reported for van der Waals junctions of TMDCs are very low, hampering the development of efficient TMDC photovoltaics.

The $V_{OC}$ is the maximum voltage produced by a photovoltaic device and, therefore, it represents the maximum energy that can be extracted from any absorbed photon under a given illumination level. If the illumination has a broadband spectrum, such as the solar spectrum, the device absorbs photons of all energies down to the minimum allowed energy, i.e. the band gap energy ($E_G$). Therefore, the so-called bandgap-voltage offset ($W_{OC} = E_G/e - V_{OC}$) is a simple measure of the fundamental energy loss in the device.[5,6] Due to thermodynamic constraints, there is a lower limit to the $W_{OC}$, which is approached when all parasitic loss mechanisms have been eliminated and only radiative recombination takes place. This minimum $W_{OC}$ value can be calculated following the Shockley-Queisser detailed balance theory[7] and, for illumination with the AM1.5G standard solar spectrum, it lies between 0.24 and 0.30 V for $E_G$ values between 1.1 and 1.9 eV (see Supporting Information, SI, for details). Devices made of epitaxial III-V materials, such as GaAs, work close to the radiative limit and exhibit the lowest experimental $W_{OC}$ with 0.31 V under the AM1.5G standard solar spectrum.[8] In the case of silicon, the record one-sun $W_{OC}$ is 0.38 V, determined by the Auger recombination inherent to this material.[9] Among emerging technologies, metal halide perovskites have reached the lowest $W_{OC}$s with 0.44 V,[10] whereas record organic solar cells exhibit higher values around 0.53 V.[11]

Solar cells based on van der Waals junctions of TMDCs reported to date, generally containing a heterojunction of $MoS_2$ and $WSe_2$, exhibit $W_{OC}$ values above 0.8 V for bulk devices and above 1 V for monolayer devices. The corresponding $V_{OC}$ values are below 0.4 V for multilayer TMDCs[3,12,13] ($E_G$ 1.2-1.3 eV) and below 0.6 V for monolayer devices[14–18] ($E_G$ 1.6-1.9 eV), even for illumination levels much higher than one-sun irradiance. Moreover, many of these devices show a pinned $V_{OC}$ [3,12,13,19], that is, a $V_{OC}$ that does not increase with illumination intensity. Interestingly, lateral junctions created on a TMDC flake by split-gating, selective chemical surface doping or using a double Schottky barrier have shown lower $W_{OC}$s than van der Waals devices, with values down to 0.54 V reported for bulk flakes[20,21] and 0.73 V for monolayers.[21,22] This suggests that the reasons behind the small $V_{OC}$ in TMDC solar cells can be found in the limitations of the van der Waals junction rather than in the materials.

In this letter, we demonstrate a $V_{OC}$ of 1.02 V and a $W_{OC}$ of 0.27 V in a TMDC-based photovoltaic cell under broadband illumination with 4 W/cm$^2$. The device consists of a vertically stacked p-$MoS_2$:Nb and n-$MoS_2$:Fe van der Waals homojunction. The main reason for the high $V_{OC}$ is the homojunction band alignment generated by the p- and n-type substitutional doping, which is more favourable than the band alignment of a heterojunction. Previous attempts at fabricating van der Waals $MoS_2$ homojunctions, however, had not shown a considerable improvement over heterojunctions;[23] the best reported result is a $V_{OC}$ pinned at 0.57 V.[24] We argue here that this can be explained by the presence of photoactive Schottky barriers at the contacts. In our devices we overcome this problem by supressing the photogeneration in these parasitic diodes, which enables the $V_{OC}$ to rise to 1 V.

Figure 1a shows the structure of the photovoltaic device, in which a p-$MoS_2$:Nb crystal (p-type, 12 nm thick) partially overlaps with an n-$MoS_2$:Fe crystal (n-type, 110 nm thick) forming a van der Waals junction with an area of 182 $\mu m^2$. The crystals were grown using the chemical transport method with a nominal doping level of 0.5%. The actual concentration of dopant atoms has been estimated from X-ray photoelectron spectroscopy (XPS) to be 0.4% Nb in the p-type crystals and 0.1% Fe in the n-type crystals. The bulk electrical doping extracted from Hall measurements is 3.0 $\times 10^{19} cm^{-3}$ in the p-material and its band diagram is unambiguously degenerate. [25] The effective electrical doping in the n-material is estimated to be ~ 7-8 $\times 10^{18}$ $cm^{-3}$. Crystal flakes were mechanically exfoliated and transferred onto a silicon wafer covered by a 90 nm thick silicon dioxide layer. Details on the fabrication procedure are available in the SI. Figure 1b shows the current density - voltage (*J*-*V*) characteristics of the device measured under broadband illumination (halogen spectrum) with three different intensities, using a two-wire configuration. The $V_{OC}$ of this device reaches 0.68 V at the highest intensity of 4 $W/cm^2$. The photoconversion efficiency is fairly constant at ~ 2%, since the improvement in $V_{OC}$ for higher light intensities is counteracted by a decrease of the fill factor (FF) associated to series resistance losses.

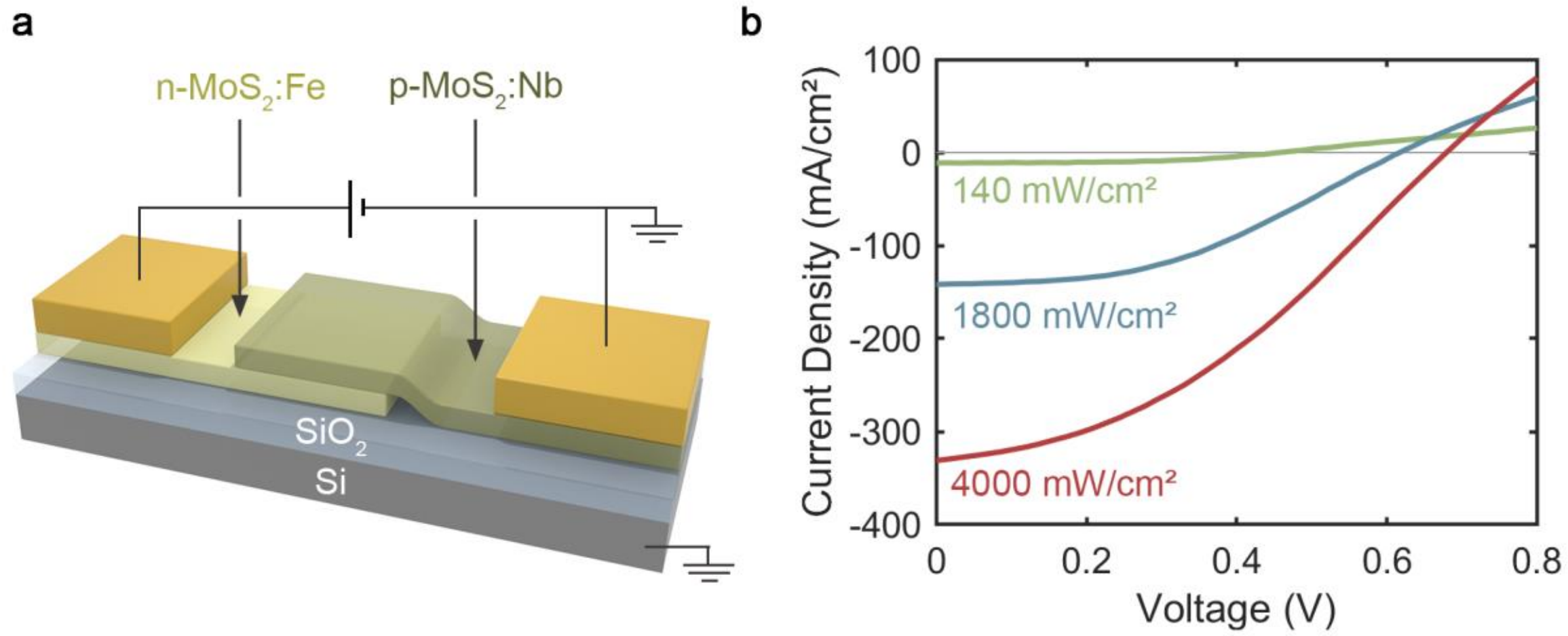

Figure 1. $MoS_2$ van der Waals homojunction solar cell on $SiO_2$/Si. a) Schematic: A p-$MoS_2$:Nb layer partially overlaps with an n-$MoS_2$:Fe layer forming a junction on an oxidized silicon substrate and they are contacted with Ni/Au electrodes. b) Two-wire *J-V* characteristics measured when the whole chip is illuminated with a broad-band (halogen) spectrum with the irradiances given by the labels.

Although the $V_{OC}$ observed in Figure 1b surpasses the highest values reported for bulk van der Waals $MoS_2$/$WSe_2$ heterojunctions (0.38 V)[3] and bulk $MoS_2$ homojunctions (0.57 V)[24], it does not reveal the real potential of the $MoS_2$ homojunction. In fact, the device can produce much higher photovoltages using the same light source and illumination density, as it is shown in Figure 2. Panels 2a to 2d show *J-V* characteristics of the device (in solid dark circles) measured in a four-wire configuration under different illumination conditions, as illustrated by the schematics. When light shines on the entire chip the device is characterised by the *J-V* curve in Figure 2a, with a $V_{OC}$ of 0.73 V. Using an iris diaphragm, it is possible to control the size of the illuminated area without modifying the illumination power density (see schematics in Figure 2a-d). If the light falling on the p-side is blocked by the diaphragm, the device is characterised by the *J-V* curve in Figure 2b with $V_{OC}$ = 0.80 V and if it is blocked the light on the n contact we find $V_{OC}$ = 0.92 V, see Figure 2c. Finally, the highest $V_{OC}$ of 1.02 V is reached in Figure 2d when both contacts are in the dark and only the pn-junction is illuminated. The different illumination curves can be explained by the presence of undesired photoactive Schottky diodes at the TMDC-metal contacts which can be modelled using a simple equivalent circuit model.

The equivalent circuit for the van der Waals $MoS_2$ homojunction solar cell is depicted in Figure 2e. The main elements are three photovoltaic generators, each one represented as a combination of a constant current generator (with photocurrent $J_L$) and a recombining diode (with dark current $J_D$). These three generators are the $MoS_2$ pn-junction, the photoactive diode at the n-Schottky contact and the photoactive diode at the p-Schottky contact. Each Schottky diode introduces a non-ohmic, illumination-dependent series resistance component in the circuit. Note that in four-wire measurements each Schottky generator appears twice. The circuit also includes a parallel resistance ($R_P$) and linear series-resistance components associated to each flake ($R_{s,n}$ and $R_{s,p}$). Solving the equation system we obtain the modelled curve $J_{model}$ (blue solid line), which is in good agreement with the experimental data $J_{exp}$ (black solid circles) in all cases in Figure 2, and also for the two-wire curves in Figure 1 and the dark curve (shown in SI, together with the set of parameters used).

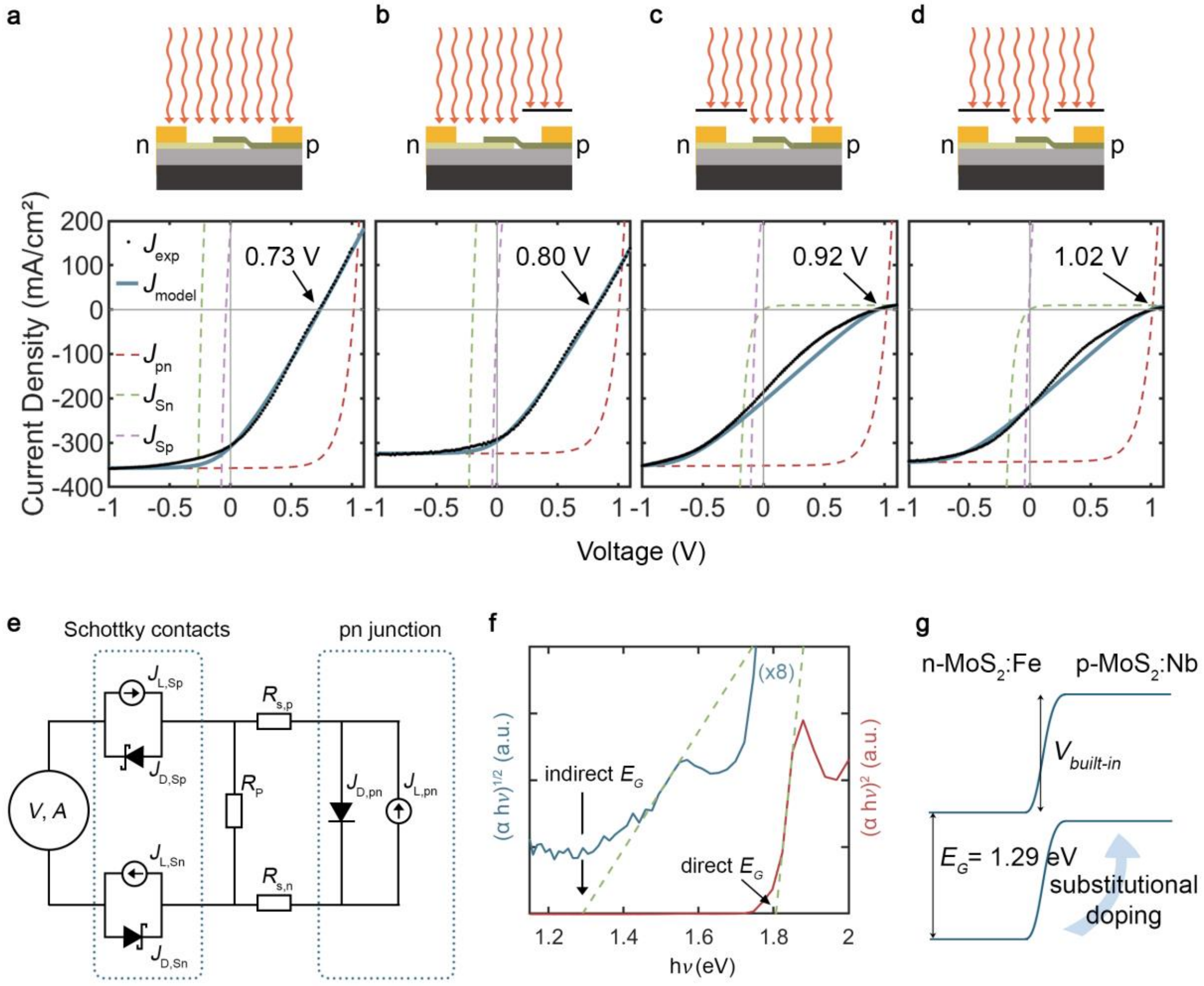

Figure 2. High photovoltage in a $MoS_2$ homojunction solar cell. a-d) Schematics and *J-V* characteristics for various illumination conditions, showing a maximum $V_{OC}$ of 1.02 V. Solid black circles are used for the experimental values and a blue line for the modelled fit. The dashed curves represent the modelled contributions from the $MoS_2$ homojunction (red) and the contacts (green and purple). The irradiance level is the same in all cases, but the illuminated region is modified using a diaphragm. e) Equivalent circuit model. f) Plot of $(\alpha h\nu)^{1/m}$ against $h\nu$ and extrapolation of the experimental $E_G$ values (with $m$=2 for indirect and $m$=1/2 for direct band gaps). g) Ideal band diagram of a $MoS_2$ homojunction with substitutional doping.

To visualize the effect of the photoactive Schottky diodes on the device performance, we have plotted in Figure 2a-d the *J-V* characteristics of each element individually. A dashed green line is used for the n-Schottky, dashed purple line for the p-Schottky, and dashed red line for the $MoS_2$ pn-junction. All measurements are consistent with the pn-homojunction generating 1.02 V.

However, because the Schottky diodes are reverse-biased with respect to the pn-junction, the photovoltage they produce under illumination is subtracted from the over-all voltage. The effect is that of a non-linear series-resistance that can not only degrade the FF but also the $V_{OC}$. Therefore, when the whole chip is illuminated (Figure 2a) both Schottky diodes together reduce the over-all $V_{OC}$ from 1.02 V to 0.73 V. The individual contributions can be decomposed by blocking selectively the light falling on either side with the diaphragm (Figure 2b and c). When both Schottky contacts are in the dark, the device $V_{OC}$ is determined solely by the one produced by the $MoS_2$ homojunction, 1.02 V. We have disregarded here the effect of shunting and linear series resistances, which are considered in the SI. Note that the voltage bias of the Schottky diodes when they are illuminated is determined by the sum of $J_{L,pn}$ and $J_{L,Sp}$ (or $J_{L,Sn}$). Therefore, in measurements where the whole device is illuminated, such as those showed in Figure 1b, a higher illumination intensity implies a larger voltage drop in the Schottky diodes, and hence a larger degradation of the FF, than what could be anticipated from the increase in $J_{L,pn}$ alone.

The behaviour of the van der Waals $MoS_2$ homojunction contacts is unconventional for solar cells. The presence of rectifying contacts in solar cells is usually associated to the appearance of a non-ohmic series resistance which reduces only the FF of the device. They can only have an effect on the $V_{OC}$ (when the current is zero) if they are photoactive, and this is not the case in conventional solar cells because, even if they are strongly rectifying, the metal covering the semiconductor keeps the Schottky barrier and its depleted region in the dark. However, this is not true for very thin TMDC-based devices. It has been found that, in this case, the depletion regions associated with Schottky barriers can extend laterally several microns away from the metal contact edge[26,27] which means that photoactive Schottky barriers are to be expected even if a metal has been evaporated on top of a TMDC. In our devices, the photocurrent generated by the Schottky contacts is consistent with depletion regions that spread ~5 µm on average from the metal edge, in good agreement with previous reports[26] (see SI for details).

The results in Figure 2a-d confirm the potential of the $MoS_2$ van der Waal homojunction as photovoltaic device, proving that it can reach $V_{OC}$ values of, at least, 1.02 V. This potential is jeopardized by the presence of Schottky barriers at the contacts and, although we can eliminate their effect on the $V_{OC}$ by keeping them in the dark and measuring with a four-wire configuration, their presence still degrades the FF of the device, and hence, its efficiency. Once the Schottky barriers are removed, the device would be characterized by the red dashed curve, which offers the prospect of reaching a photoconversion efficiency of ~6 % in this simple unoptimized device structure, rendering the use of $MoS_2$ a promising approach to TMDC based photovoltaics. However, eliminating the Schottky barriers completely at TMDC/metal interfaces is not a simple task and will require further technological developments, as pointed out before.[27,28] The comparison of Figure 2b and 2c shows that, in our device, the n-Schottky barrier is more rectifying than the p-Schottky barrier. In broad terms, this is to be expected from the Schottky-Mott rule, because the work function of Ni (-5.2 eV) is closer to the valence band maximum energy of bulk $MoS_2$ (-5.5 eV) than to the conduction band minimum energy (-4.2 eV). Still, the real height of the Schottky barrier cannot be known from the work functions alone because the metal evaporation triggers band structure changes in TMDC and leads to Fermi-level pinning,[21,27,29–32] This implies that to achieve ohmic contacts to TMDCs it is not sufficient to evaporate metals with well-aligned work functions. Therefore, more sophisticated strategies will be required to unlock the photoconversion potential of the $MoS_2$ homojunction in practical devices.

Once we have established the $V_{OC}$ of the homojunction, the value of the band gap $E_G$ is critical for the discussion of the $W_{OC}$. It is known that bulk $MoS_2$ exhibits two bandgap energies: an indirect transition at around 1.3 eV and a direct transition at a higher energy. When the $MoS_2$ sample is thinned, the indirect transition is shifted upwards in energy and, in the ML limit, the direct band gap dominates the opto-electronic properties[1]. To determine $E_G$ in our samples we apply the generic relationship[33] between the absorption coefficient ($\alpha$) of a semiconductor, estimated in this case from the external quantum efficiency (EQE), and the photon energy ($h\nu$), obtaining the plot in Figure 2f (see SI for details). We find an indirect band gap with $E_G$ = 1.29 ± 0.04 eV and a direct one with $E_G$ = 1.82 ± 0.02 eV in agreement with the reported values for bulk $MoS_2$.[1] In solar cells where the absorber material has a smaller indirect band gap and a larger direct band gap, such as the ubiquitous silicon solar cells, the smallest of the two limits the $V_{OC}$.[34] In this case we find $W_{OC}$ = 0.27 V at an illumination intensity approximately equivalent to 40 suns. At this illumination intensity, the radiative $W_{OC}$ for an ideal device which absorbs all photons above $E_G$ = 1.29 eV is 0.16 V (see SI for calculation applying Shockley-Queisser's detailed balance theory[7]). Therefore, our experimental result is encouragingly close to the radiative limit.

The band diagram in Figure 2g illustrates why approaching the radiative limit established by the detailed balance theory is possible in a homojunction and not in a heterojunction. In Figure 2g we draw an idealised band diagram (disregarding possible surface bending effects) of a bulk $MoS_2$ homojunction. Doping, and in particular degenerate doping as in our material,[25] causes the bands to be shifted so far that the built-in voltage (times the electron charge *e*) approaches the band gap. With suitable contacts and minimal non-radiative recombination, the $V_{OC}$ approaches the built-in voltage. Therefore, in a homojunction the $V_{OC}$ is ultimately limited by $E_G$, which is not the case of a heterojunction. For example, intrinsic bulk $MoS_2$ in combination with intrinsic bulk $WSe_2$ results in a band diagram in which the alignment is given by the respective electron affinities (band diagrams of $MoS_2$ / $WSe_2$ heterojunctions are shown in the SI). If a forward voltage bias is applied, one band becomes flat when the bias equals the smaller of the two band offsets, in this case 0.4 V. Therefore, the smaller band offset limits the $V_{OC}$ in a heterojunction. This barely improves if the device consists of MLs, although their band gap energies are much larger (also shown in the SI). The smaller of the two offsets at the band edges, i.e. the ultimate limit to the $V_{OC}$, in an intrinsic ML $MoS_2$ / ML $WSe_2$ union is 0.7 eV.

Having discussed the $V_{OC}$, in the following we focus on the photocurrent generated by the homojunction solar cell, which is determined by the absorbance. We calculate the light energy flux inside of a $MoS_2$/substrate structure applying the generalized matrix method as described by E. Centurioni[35] and using literature values of the optical constants[36]. Figure 3a shows the calculated absorbance, transmittance, and reflectance of a $MoS_2$ slab for varying thickness *d* from 1 ML to 200 nm. The values are representative of photon energies above the direct optical $E_G$ (1.82 eV) because photons between the two band gaps of $MoS_2$ are only weakly absorbed in an ultra-thin device. The plotted values correspond to the fraction of photons from the AM1.5G spectrum that are absorbed, transmitted, or reflected in the range 300-680 nm. Three cases are presented: a free-standing $MoS_2$ slab (solid lines), deposited onto a glass substrate (dashed lines) and deposited onto a Si substrate with a 90 nm thick $SiO_2$ layer on top (dotted lines).

For free-standing $MoS_2$, the absorbance plotted in Figure 3a has a local maximum of 38% at $d$ = 7 nm and an absolute maximum at $d$ = 61 nm of 51%. Above $d$ = 100 nm the transmittance is negligible, and both the absorbance and the reflectance reach virtually constant values close to 50%. The dotted and dashed lines in Figure 3a, representing devices on a substrate, show a similar trend and only differ slightly from free-standing $MoS_2$ due to the different optical environments and the reflections at the additional interfaces. The difference to bare $MoS_2$ is greatest at small thicknesses and decreases monotonically with increasing thickness up to $d$ = 100 nm. Above that point, their absorbance is also constant and close to 50%. We identify two regions of interest for photovoltaic applications. Extremely thin devices between 1 ML and 10 nm can render high efficiency due to low reflectance losses, especially for semi-transparent applications (e.g. power-generating windows) or if light-trapping techniques are used to minimise transmission losses. To this end, it still has to be demonstrated that van der Waals homojunctions with ML or few ML thickness can be produced via substitutional doping. Recent reports on stable conductivity of $MoS_2$:Nb in the ML limit are encouraging in this respect.[37] The second region of interest is the range around 100 nm thickness, which is exploited in this work. In this range half of the incoming power is lost by reflection due to the high refractive index of $MoS_2$ compared to other semiconductors commonly used in photovoltaics, such as GaAs or Si. Therefore, it can be concluded that the photoconversion efficiency of these devices could be approximately doubled with the design and application of an optimised anti-reflection coating.

Figure 3b shows an example of the calculated absorbance spectra for a $MoS_2$ flake on glass with $d$ = 2, 15, 50 and 120 nm. The spectra show the two excitonic peaks A and B characteristic of $MoS_2$, at 660±5 and 600±5 nm respectively, corresponding to the absorption by direct transitions at the K point of the Brillouin zone, with an energy difference given by the valence band splitting due to spin–orbit and interlayer coupling.[1,38] In some cases, additional peaks appear that are related to multiple reflection and interference in the semiconductor layer. For example, in the red curve a peak is observed at λ = 530 nm for $d$ = 50 nm, for which the resonance condition $d = j\lambda/4\mathrm{n}$ is fulfilled, where $j$ is an integer and $n$ is the refractive index.[39]

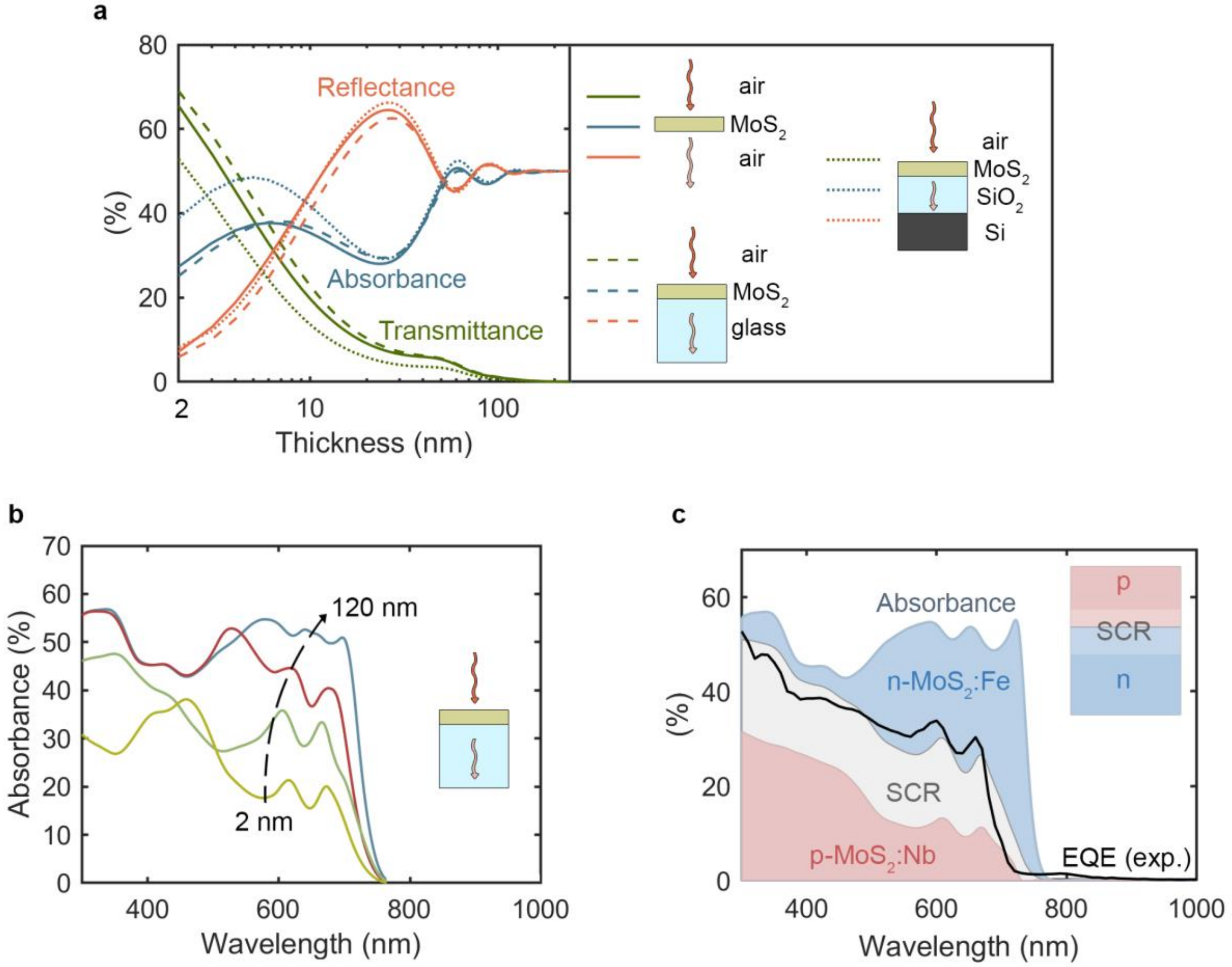

Figure 3. Photocurrent in a $MoS_2$ homojunction solar cell. a) Calculated absorbance, reflectance and transmittance (weighted average over the AM1.5G solar spectrum) for varying thickness of $MoS_2$, either free-standing, on $SiO_2$ (90 nm)/Si, or on glass, showing that a maximum of absorbance is reached for a layer thickness above ~100 nm. b) Calculated absorption spectra of $MoS_2$ on glass for thicknesses $d$ = 2, 15, 50 and 120 nm. c) Calculated absorption spectra for our $MoS_2$ on glass device and comparison to experimental EQE data.

Figure 3c compares an experimental EQE measured on a $MoS_2$ van der Waals homojunction with the calculated absorbance of a $MoS_2$ slab of the same thickness on the same type of substrate. For this measurement we used a device fabricated on glass (12 nm p-$MoS_2$:Nb on 262 nm n-$MoS_2$:Fe, area 293 µm²). This device was made on glass instead of oxidized silicon because the EQE measurement is performed using a lock-in amplifier and a chopped light source, and therefore, implies the measurement of an alternating current. When a $SiO_2$/Si substrate is used the EQE measurement is distorted because the $SiO_2$ layer acts as a capacitor that allows the photogeneration in the Si substrate to alter the measurement. The performance of the $MoS_2$ on glass device is inferior to the one on oxidize silicon due to surface doping of the flakes (see SI for experimental and modelled *J-V* curves). Therefore, the EQE plotted in Figure 3c can be considered a lower bound to the EQE of the $MoS_2$ van der Waals homojunctions.

The calculated absorbance in Figure 3c is broken down into three contributions from the three electrical regions of the device: space charge region (SCR, 16.5 nm), neutral part of the p-flake (3.5 nm) and neutral part of the n-flake (249 nm). We have calculated the SCR thickness using the total depletion approximation[40] and the doping levels $3.0 \times 10^{19}$ and $7.5 \times 10^{18}$ $cm^{-3}$ estimated from the characterisation of doped bulk $MoS_2$ samples. It is important to note that the measured EQE accounts approximately for the sum of the p-flake and the SCR contributions, and the neutral part of the n-flake does not contribute a significant amount to the photocurrent. This points to a higher quality (larger minority carrier diffusion length) of the p-doped material over the n-doped material (note that under short-circuit conditions all carriers photogenerated in the SCR are collected, independently of the material quality). The integration of the EQE over the AM1.5G solar spectrum results in a total short-circuit current of 6.6 $mA/cm^2$. To estimate the room for improvement of this value, it can be compared to the integration of the theoretical absorbance over the same spectrum, which is 11.8 $mA/cm^2$. This yields an internal quantum efficiency (IQE) of 56 % weighted over the range of interest. To increase the IQE, the collection from the n-flake needs to be improved. As pointed out before, the absorbance, and therefore the photocurrent, could be almost doubled if a suitable anti-reflection coating for this material is developed.

To test the radiative efficiency of the doped $MoS_2$ material we have fabricated light emitting tunnelling devices. These devices consist of a $MoS_2$ flake on which a thin hexagonal boron nitride (h-BN) layer (1-3 ML) and few-layer graphene (FLG) have been deposited (see Figure 4a). This contact is quasi-transparent and allows homogenous charge injection over a large area, producing a 2D junction which enhances the collection of EL from TMDC-based devices.[41] The $MoS_2$ flake is deposited on top of a thick h-BN supporting layer and a second contact is implemented laterally using FLG. Figure 4b shows the current-voltage (*I-V*) characteristic of a device containing a bulk p-$MoS_2$:Nb flake. The *I-V* curve is exponential in both quadrants, revealing that the electrical behaviour of the device is dominated by tunnelling through the h-BN. An idealised band diagram of the tunnelling device is depicted in Figure 4c.

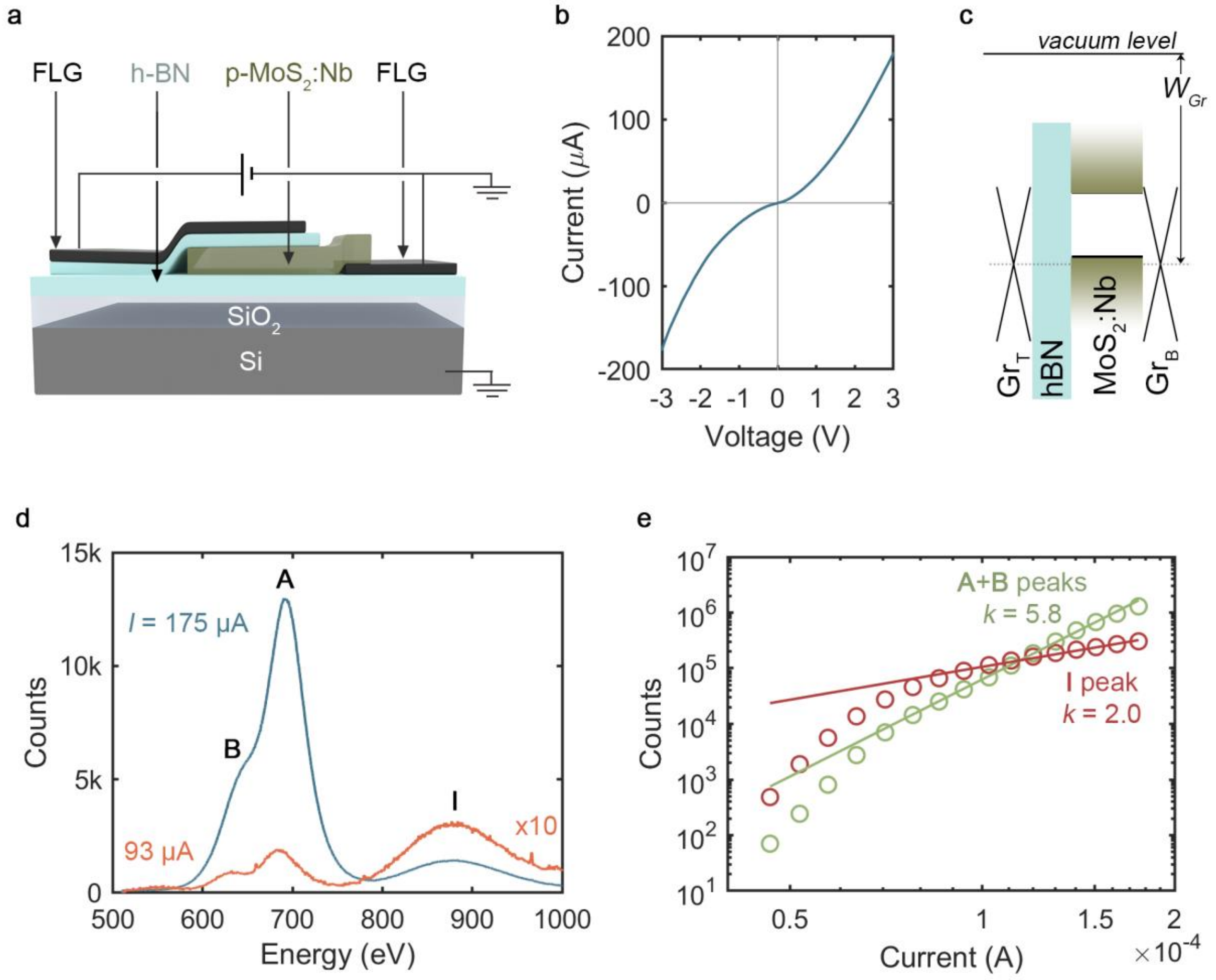

Figure 4. EL from FLG/BN/p-$MoS_2$:Nb tunneling structure. a) Schematic of the device. A bulk p-$MoS_2$:Nb flake is contacted from above by a FLG/BN tunneling contact, and laterally by FLG. b) Current-voltage curve. d) EL spectra at injection currents *I* = 93 µA (red) and 175 µA (blue), showing the exciton peaks **A** and **B** as well as the indirect transition **I**. e) Log-log plot of the EL signal (with background subtracted) integrated between 580 and 780 nm and between 780 and 1000 nm, vs current.

We find that the device produces EL for bias voltages above 1.6 V. Figure 4d shows the EL spectrum when 93 µA (red curve) and 175 µA (blue curve) are injected. The exciton peaks A and B are at 684±1 nm and 622±2 nm. The redshift of the exciton peaks by ~70 meV compared to the absorption spectra can be attributed to the interaction of the $MoS_2$ surface with the different surrounding media, glass and h-BN, in agreement with previous studies on luminescence in $MoS_2$.[42,43] Remarkably, EL from the indirect transition I at 880±5 nm is also observed.

As we have discussed above, the optical properties of $MoS_2$ in bulk form are dominated by the indirect band gap at 1.3 eV. Only when the thickness is decreased, approaching 1 ML, does the band gap shift upwards in energy, leading to a crossover to the direct band gap with energy 1.8 eV. For this reason, photoluminescence spectra from ML $MoS_2$ are four orders of magnitude stronger than from bulk samples,[1] and the EL, which is strong in MLs, can generally not be detected in bulk devices.[44] EL from multilayer intrinsic $MoS_2$ has been demonstrated only under strong electric fields and has been attributed to electric-field-induced carrier redistribution from the lowest energy points (indirect bandgap) to higher energy points (direct bandgap) in k-space.[45] Our observation of EL from a bulk $MoS_2$:Nb highlights the potential of substitutional doping of $MoS_2$ for the fabrication of opto-electronic devices and it also sheds light on why the $W_{OC}$ exhibited by $MoS_2$ homojunction solar cells is so close to the radiative $W_{OC}$ limit in spite of the indirect nature of bulk $MoS_2$. A log-log plot of the integrated EL emission versus the injected current (Figure 4e) reveals that the recombination is only partially radiative, as is expected given the nature of the transitions (a fully radiative device would exhibit a slope $k$ =1). Similar devices fabricated with n-doped or intrinsic $MoS_2$ did not produce any measurable EL. This indicates a lower radiative efficiency in the n-doped material, which is consistent with our analysis of the EQE from $MoS_2$ homojunctions, where we determined that the n-flake contributes much less to the photocurrent than the p-flake.

In conclusion, we report a $MoS_2$ homojunction produced through substitutional doping that exhibits the highest yet-reported $V_{OC}$ in a photovoltaic van der Waals TMDC structure, 1.02 V under broadband illumination with 4 W/cm$^2$. This corresponds to a $W_{OC}$ value of 0.27 V, to be compared to 0.16 V calculated for an ideal, fully radiative device under the same illumination. The generation of a photovoltage close to the radiative limit is consistent with the observation of electroluminescence from tunnelling structures containing p-doped $MoS_2$, which is a remarkable finding given the indirect nature of bulk $MoS_2$. We find that in our devices the high $V_{OC}$ produced by the $MoS_2$ homojunction can only be observed if the photogeneration in the Schottky barriers present at the TMDC/metal interfaces is avoided. This result confirms the potential of TMDCs, and in particular substitutionally doped $MoS_2$, for the development of highly efficient, ultra-thin and ultralight weight photovoltaic devices. From the analysis of the *J-V* curves and the EQE we outline the next steps to realize high efficiency, which are the elimination of the Schottky barriers at the metallic contacts, the implementation of an anti-reflection coating and the improvement of the radiative efficiency of the n-doped flake.

ACKNOWLEDGMENTS

This work has been funded by the Spanish Science Ministry under Grant TEC2017-92424-EXP and by the Fundación Ramón Areces within the research project SuGaR. D.Y.L. acknowledges financial support from the Ministry of Science and Technology of Taiwan under grand no. MOST 108-2221-E-018-010. C.B-B and M.H.Z. are grateful to the Universidad Politécnica de Madrid for funding through the Predoctoral Grant Programme. S.A.S acknowledges a Juan de la Cierva Fellowship (FJC2018-036517-I) and E.A. acknowledges a Ramón y Cajal Fellowship (RYC-2015-18539), both funded by the Spanish Science Ministry. K.W. and T.T. acknowledge support from the Elemental Strategy Initiative conducted by the MEXT, Japan, Grant Number JPMXP0112101001, JSPS KAKENHI Grant Number JP20H00354 and the CREST(JPMJCR15F3), JST. Additional financial support was provided by the Leverhulme Trust [grant number RPG-2016-104]. P.H.B thanks the Leverhulme Trust for the award of a Research Fellowship [RF-2019-460].

COMPETING INTERESTS

The authors declare no competing interests.

SUPPORTING INFORMATION

Detailed balance calculations. Device fabrication. Device characterisation. Equivalent circuit model and parameters. Supplementary experimental and modelled *J-V* curves. Band gap energy determination. Device micrographs and AFM profiles. Supplementary band diagrams. Comparison to other works.

TOC GRAPHIC

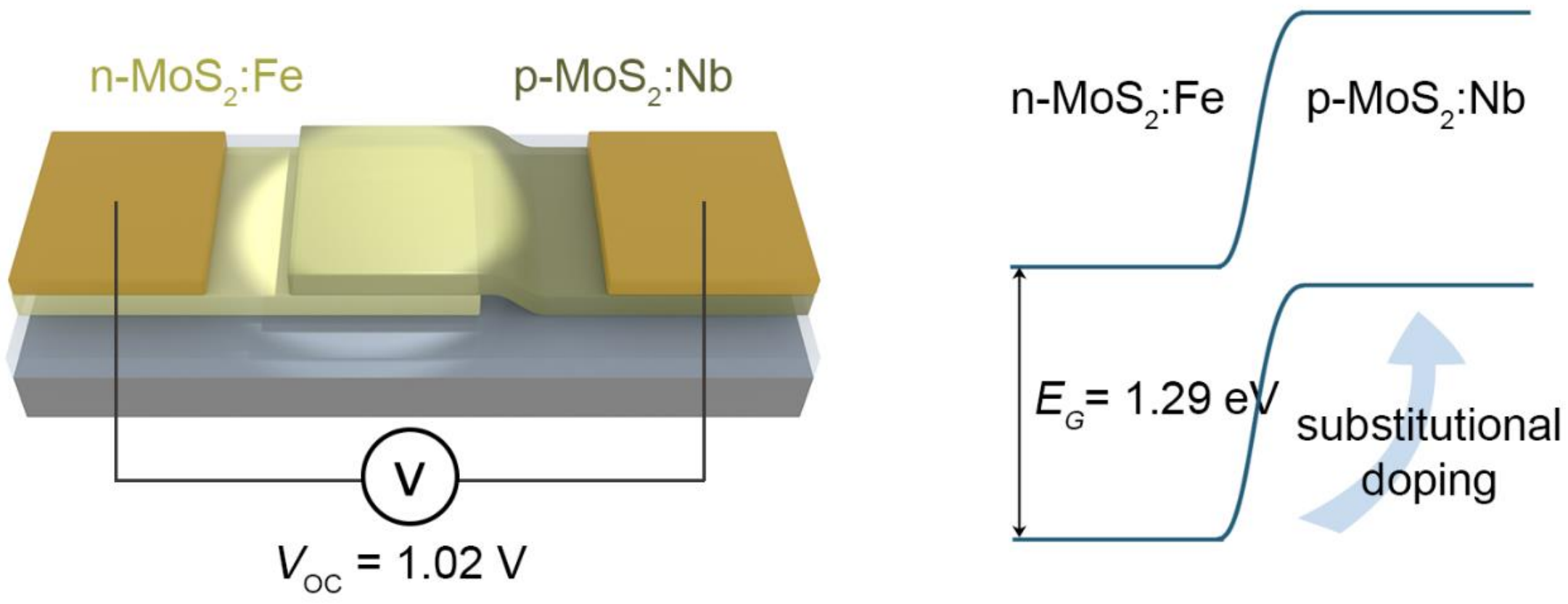

## Supporting Information

# High open-circuit voltage in transition metal dichalcogenide solar cells

Simon A. Svatek[1], Carlos Bueno-Blanco[1], Der-Yuh Lin[2], James Kerfoot[3], Carlos Macías[1], Marius H. Zehender[1], Ignacio Tobías[1], Pablo García-Linares[1], Takashi Taniguchi[4], Kenji Watanabe[5], Peter Beton[3], Elisa Antolín[1]*

1. Instituto de Energía Solar, Universidad Politécnica de Madrid, Avenida Complutense 30, 28040 Madrid, Spain

2. Department of Electronics Engineering, National Changhua University of Education, Changhua 50007, Taiwan

3. School of Physics and Astronomy, University of Nottingham, Nottingham NG7 2RD, U.K.

4. International Center for Materials Nanoarchitectonics, National Institute for Materials Science, 1-1 Namiki, Tsukuba 305-0044, Japan

5. Research Center for Functional Materials, National Institute for Materials Science, 1-1 Namiki, Tsukuba 305-0044, Japan

*e-mail: elisa.antolin@upm.es

**SI_Section 1. Detailed balance calculations.**

We use the detailed balance formalism to calculate the maximum $V_{OC}$ that can be produced by an ideal single-junction solar cell with a given band gap energy, $E_G$, and under a given illumination density. The original reference for the theoretical framework is Ref. 1 and a comprehensive explanation can be found in Ref. 2.

The maximum $V_{OC}$ is produced by an ideal device in which the absorptivity is 1 (it absorbs all photons with energy above $E_G$) and radiative recombination is the only loss mechanism. The current density ($J$, mA/cm$^2$) of such a device as function of the bias voltage ($V$) is given by

$$J(V) = J_\mathrm{L} - J_\mathrm{D}(V) \qquad (1)$$

Therefore, the radiative $V_{OC}$ can be extracted from

$$J_\mathrm{L} - J_\mathrm{D}(V_\mathrm{OC}) = 0 \qquad (2)$$

The illumination current, $J_L$, is given by

$$J_\mathrm{L} = e \int_{E_\mathrm{G}}^{\infty} F(E)\, dE \qquad (3)$$

where $e$ is the electron charge and $F$ (expressed in s$^{-1}$ cm$^{-2}$ eV$^{-1}$) is the number of photons in an energy interval d$E$ of the AM1.5G standard solar spectrum[3], per unit of time and area.

The dark current, $J_D$, is given by

$$J_\mathrm{D}(V) = e\, \frac{2\pi}{h^3 c^2} \int_{E_G}^{\infty} \frac{E^2}{\exp\left(\frac{E-eV}{kT}\right)-1}\, dE \qquad (4)$$

where $h$ is Planck's constant, $c$ the speed of light in vacuum, $k$ Boltzmann's constant and $T$ the device temperature (here 300 K). Note that we have considered the most ideal case, that is, a solar cell with an ideal back mirror which only emits photons through the front surface. For a free-standing semiconductor slab, which emits photons at the same rate through the front and rear surfaces, the dark current in equation (4) would be multiplied by a factor 2, and the radiative $V_{OC}$ would be reduced accordingly.

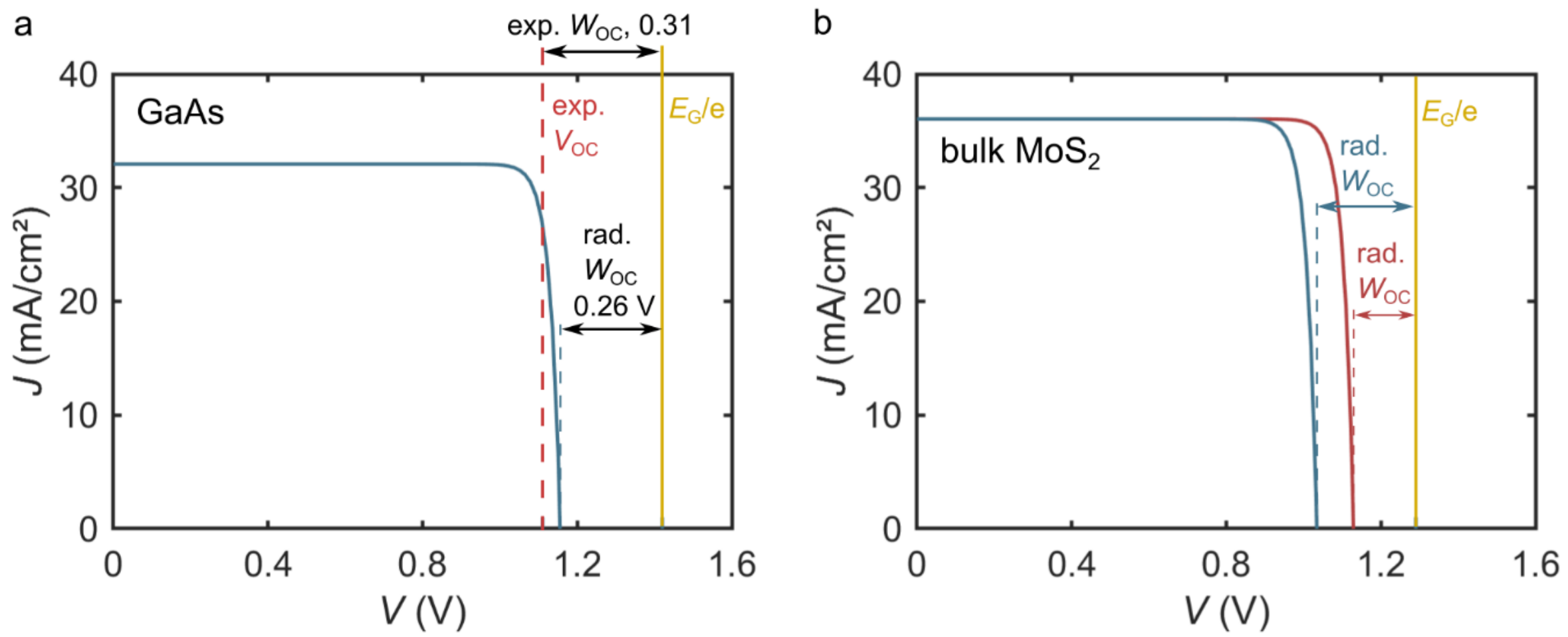

Figure S1. a) *J-V* curve of a GaAs device working in the radiative limit, calculated using Eq. (1) to (4), and compared it to the best reported experimental $V_{OC}$. b) Same calculation for a device made of bulk $MoS_2$ (blue curve). We also include the calculation for an illumination of 40 AM1.5G suns (red curve), which is eqivalent to the one used in our experimental set-up. It is obtained multiplying $J_L$ (eq. 3) by 40. In the plot $J$ has been scaled down, dividing it by 40, to faciliate comparison.

## SI_Section 2. Device Fabrication.

An exhaustive study of the material composition and growth of $MoS_2$:Fe and $MoS_2$:Nb is given in Suh et al.[4] and Wang et al.[5]. The crystals were exfoliated onto $SiO_2$ (85 nm)/Si wafers (Siltronix) or glass using adhesive tape (BT-150E-CM, Nitto). To assemble the material stack we prepared polymer stamps, consisting of a glass slide onto which a drop of Sylgard® 184 polydimethylsiloxane covered with polypropylene carbonate (15 % by weight in anisole, Merck) is deposited. The top material flake was picked up by bringing the stamp into contact with the flake on the substrate which is heated to 45 °C. The stamp is then retracted and removes the flake from the substrate. This pick-up is repeated to form the p-n junction. To release the p-n junction the stamp is brought into contact with another substrate and heated to 90 °C, which causes the PPC to melt and the p-n junction to stay on the substrate. This method warranties that the interfaces forming the p-n junction have not been in contact with any polymer. Residual PPC from the upper-most surface is washed off using chloroform. Metal contacts (40 nm Ni/70 nm

Au) are deposited using thermal evaporation in combination with optical lithography and lift-off. For the device fabricated on glass, the metal contacts where pre-patterned and the semiconductor flakes were deposited on top of them. The FLG/BN tunnelling structures have been fabricated analogously.

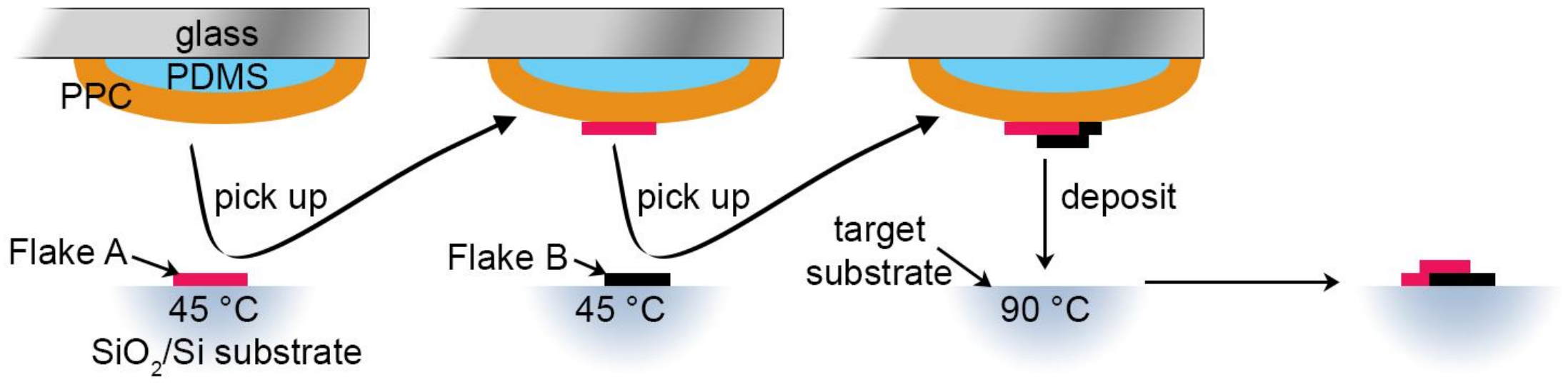

Figure S2: Schematic showing the technique to pick-up and release flakes.

**SI_Section 3. Device Characterisation.**

All measurements were aquired in atmosphere at room temperature. *J-V* characteristics were measured using a sourcemeter (either Keithley 2400 or Keysight B2901A). The illumination source for *J-V* curves was a halogen lamp coupled to a microscope (Motic BA310 MET-H). The iris diaphragm of the microscope allowed us to partially block the light impigning on certain regions of the chip. The light intensity was adjusted to 4000 mW/cm² using a calibrated Si solar cell. Quantum efficiency measurements were carried out using an Oriel Cornerstone 260 Monochromator equipped with order sorting filters, a quartz tungsten halogen lamp, a low-current transimpedance pre-amplifier (Stanford Research SR570 DSP), a mechanical chopper and a Stanford Research SR830 DSP Lock-In Amplifier. The photocurrent was calibrated against a NIST-traceable calibrated Si photodetector (Newport STPVCERT) and then it was scaled to obtain the absolute quantum efficiency using a LED-based solar simulator (WaveLabs LS2). The atomic force microscopy (AFM) was done using a Multimode Nanoscope III A in tapping mode. To measure electroluminescence we used a Horiba MicOS optical spectrometer with a 50x objective, NA: 0.5, and 150 l/mm grating.

**SI_Section 4. Equivalent circuit model.**

The model to fit the *J-V* curves of the $MoS_2$ van der Waals homojunction device is based on the equivalent circuit depicted in Fig. S3a. For 4-wire measurements we use the circuit in Fig. S3b.

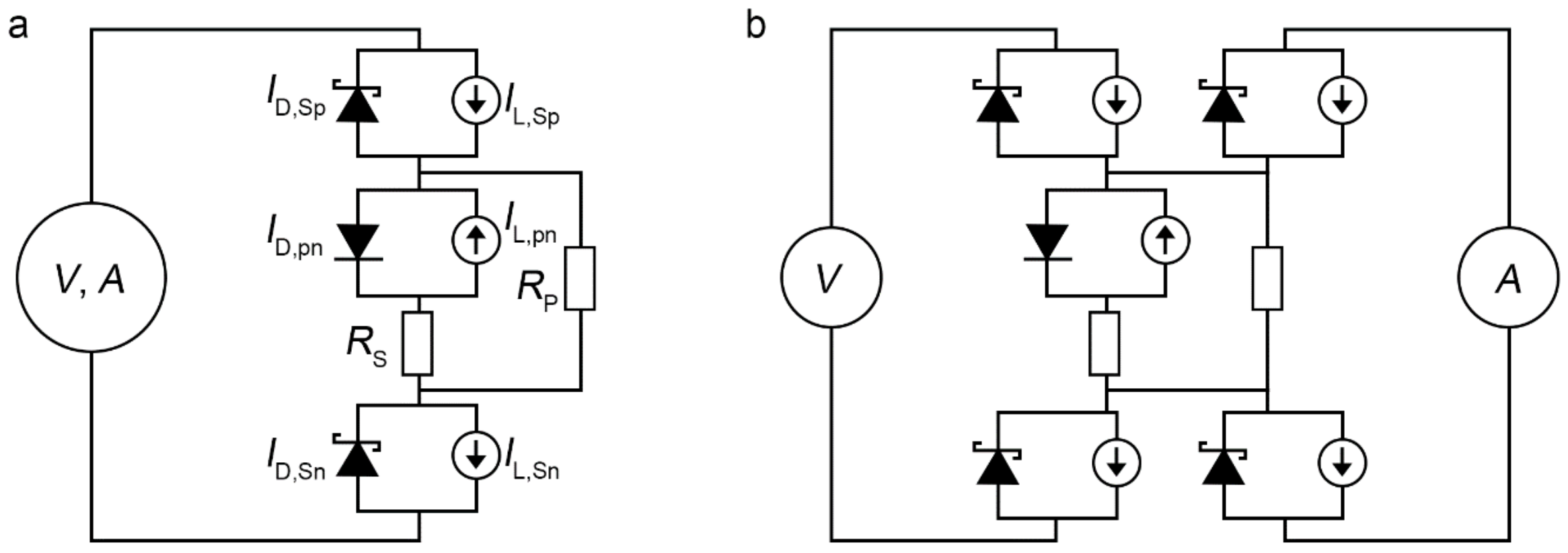

Figure S3: Equivalent circuit model of the $MoS_2$ homojunctions when measured in 2-wire (a) and 4-wire (b) configuration. The n-$MoS_2$-p-$MoS_2$ junction generates under illumination a current $I_{L,PN}$. The Schottky contacts generate, when illuminated, a current $I_{L,Sp}$ and $I_{L,Sn}$.

There are three photovoltaic elements, each represented in the circuit as a combination of a constant current generator (photogeneration) and a diode (recombination): the $MoS_2$ homojunction, which generates a photocurrent $I_{\mathrm{L,pn}}$ ; the Schottky at the n-contact, which generates a photocurrent $I_{\mathrm{L,Sn}}$ ; and the Schottky at the p-contact, which generates a photocurrent $I_{\mathrm{L,Sp}}$. The respective dark currents of the diodes are $I_{\mathrm{D,pn}}$, $I_{\mathrm{D,Sn}}$ and $I_{\mathrm{D,Sp}}$.

The three dark currents can be modelled as:

$$I_{D,\mathrm{pn}} = I_{0,\mathrm{pn}} \exp\left(\frac{qV}{n_{\mathrm{pn}}kT} - 1\right) \tag{5}$$

$$I_{\mathrm{D,Sn}} = I_{0,\mathrm{Sn}} \exp\left(\frac{qV}{n_{\mathrm{Sn}}kT} - 1\right) \tag{6}$$

$$I_{\mathrm{D,Sp}} = I_{0,\mathrm{Sp}} \exp\left(\frac{qV}{n_{\mathrm{Sp}}kT} - 1\right) \tag{7}$$

where $I_{0,XX}$ is the reverse saturation current of the respective diode, $n_{XX}$ the ideality factor, $q$ the electron charge, $k$ the Boltzmann constant and $T$ the temperature.

$R_P$ is the shunt resistance of the device, which is constant for all measurements on that device. $R_S$ is the linear series resistance introduced by the n- and p-flake in the transport of current to the metal contacts. $R_S$ is strongly illumination dependent, decreasing when the regions around the junction and the metal contacts are illuminated. Since the n- and p-flakes are degenerately doped, we attribute this illumination dependence to the existence of depleted regions (associated to the lateral extension of the depleted regions of the contacts and to surface doping effects cased by the contamination and by the substrate properties). Therefore, when the flakes are in the dark, their series resistance increases. This effect by itself would result in an improvement of the FF with the illumination intensity. However, in our device we observe the opposite tendence, the FF degrades as the illumination power is increased (for example in Figure 1b) because the voltage drops generated at the Schottky diodes become larger, counteracting the improvement in the linear series-resistance. Note that the series resistive effects on the FF in this device, including the linear component and especially the non-linear component, are strong enough to affect the measured $J_{SC}$ value.

It is important to note that the interplay of $R_P$ and $R_S$ in our circuit model enables that the $V_{OC}$ is reduced by the effect of $R_S$. This is not the case in simple solar cell circuits, where $R_S$ is in series with the current meter and, if the measured current is zero, the voltage drop over $R_S$ is also zero. Here $R_P$ and $R_S$ are interconnected in a way that $R_S$ can induce a reduction in the photovoltage, even when measured under open-circuit conditions. If we define $V_{\mathrm{pn}}^{*}$ as the voltage drop across the junction itself, then

$$V_{\mathrm{pn}} = V_{\mathrm{pn}}^{*} + \left(I_{\mathrm{L,pn}} - I_{\mathrm{D,pn}}\right) \cdot R_{\mathrm{S}} \quad (8)$$

is the resistively-reduced junction voltage. The second term is negative because $I_{L,PN}$ is negative. $V_{PN}$ equals also the voltage across $R_P$. Although the difference between $V_{\mathrm{pn}}^{*}$ and $V_{\mathrm{pn}}$ is very small in actual devices, our experimental curves cannot be fitted consistently if we use a different interconnection between $R_S$ and $R_P$. The measured $V_{OC}$ is then:

$$V_{OC} = V_{pn} - V_{Sn} - V_{Sp} \tag{9}$$

Table S1 summarizes the parameters that have been used to fit the *J-V* curves shown in Figure 2 in the main text. Note that the small variations in short-circuit current in the experimental curves are caused by slight differences in the amount of stray light when the diaphragm is position modified. In Table S1 we have converted the current values into current density values using the overlapping area of each junction. This is an approximation because, as discussed in the main text, the real junction areas can spread beyond the overlapping area. Also, when a bias is applied, the space charge region (SCR) spreads or shrinks vertically and/or laterally and this has an impact on the amount of photocurrent and dark current because, in thin devices made of TMDCs, unlike conventional devices, the volume of the SCR represents a large portion of the device. Therefore, a exact modelling of the device would require the introduction of voltage-dependent $I_L$ and $I_0$ parameters. Nevertheless, the good areement between experiment and model in Figure 2 indicates that using the approximation of constant junction area and constant SCR volume is valid, although it results in a high apparent ideality factor for the pn-junction.

Table S1: Modelling parameters used for Figure 2

| | |
|---|---|
| $R_p$ (Ωcm$^2$) | 2093 |
| $R_s$ (Ωcm$^2$) | from 74.6 (in the dark)<br>to 0.5 (when illuminated with 4W/cm$^2$) |
| $n_{np}$ | 3.8 |
| $n_{Sn}$ | 2.0 |
| $n_{Sp}$ | 2.0 |
| $J_{0,np}$ (mA/cm$^2$) | 0.012 |
| $J_{0.Sn}$ (mA/cm$^2$) | 9 |
| $J_{0,Sp}$ (mA/cm$^2$) | 327 |
| $A_{np}$ (µm$^2$) | 182 |
| $A_{Sn}$ (µm$^2$, under metal) | 216 |
| $A_{Sp}$ (µm$^2$, under metal) | 95 |
| $J_{L.Sn}$ / $J_{L.pn}$ | 1.65 |
| $J_{L.Sp}$ / $J_{L.pn}$ | 0.7 |

All curves in Figure 2 are fitted using a constant ratio between the photocurrent produced by each Schottky contact and the photoccurent produced by the pn-junction. From this ratio we can estimate an average width of the $MoS_2$ region that produces photocurrent around the metal edge. Assuming that the photocurrent density in the Schottkys is similar to the pn-junction, the width of the photogenerating region can be estimated to ~6 µm around the n contact and ~4 µm around the p-contact.

**SI_Section 5. Experimental and modelled two-wire *J-V* curves.**

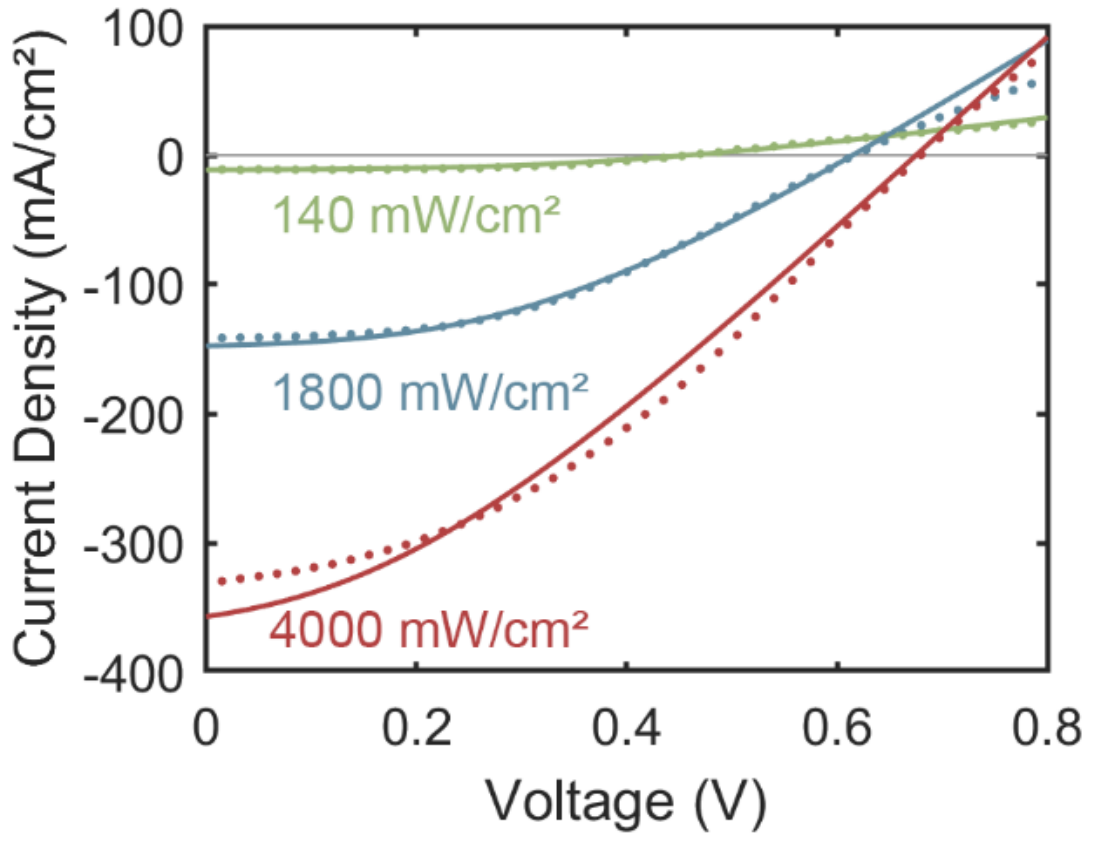

Figure S4: Experimental two-wire *J-V* characteristics of the device on $SiO_2$/Si shown in Fig. 1b in the main text (solid circles) together with corresponding fitted curves (solid lines) using the model from Figure S3a.

## SI_Section 6. Experimental and modelled two-wire *J-V* curves of the device on glass

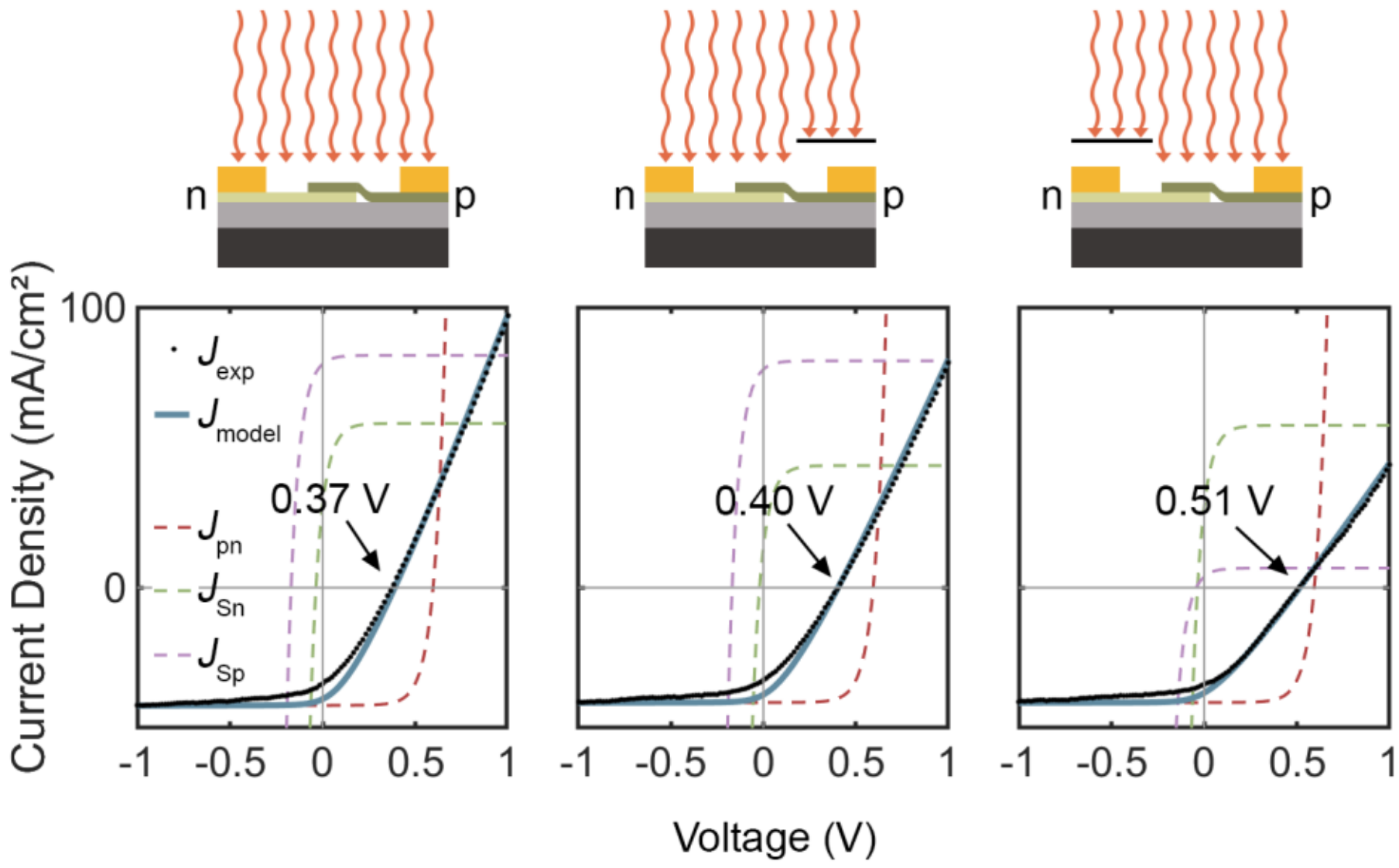

Figure S5: Schematics and *J-V* characteristics of the device on glass for various illumination conditions (the device used for EQE measurements). The light source is a halogen lamp with intensity approximately 600 mW/cm². Blocking the light on both sides was not possible due to the smaller device size. The modelled curves have been calculated using the model from Figure S3a.

## SI_Section 7. Experimental and modelled dark *J-V* curves.

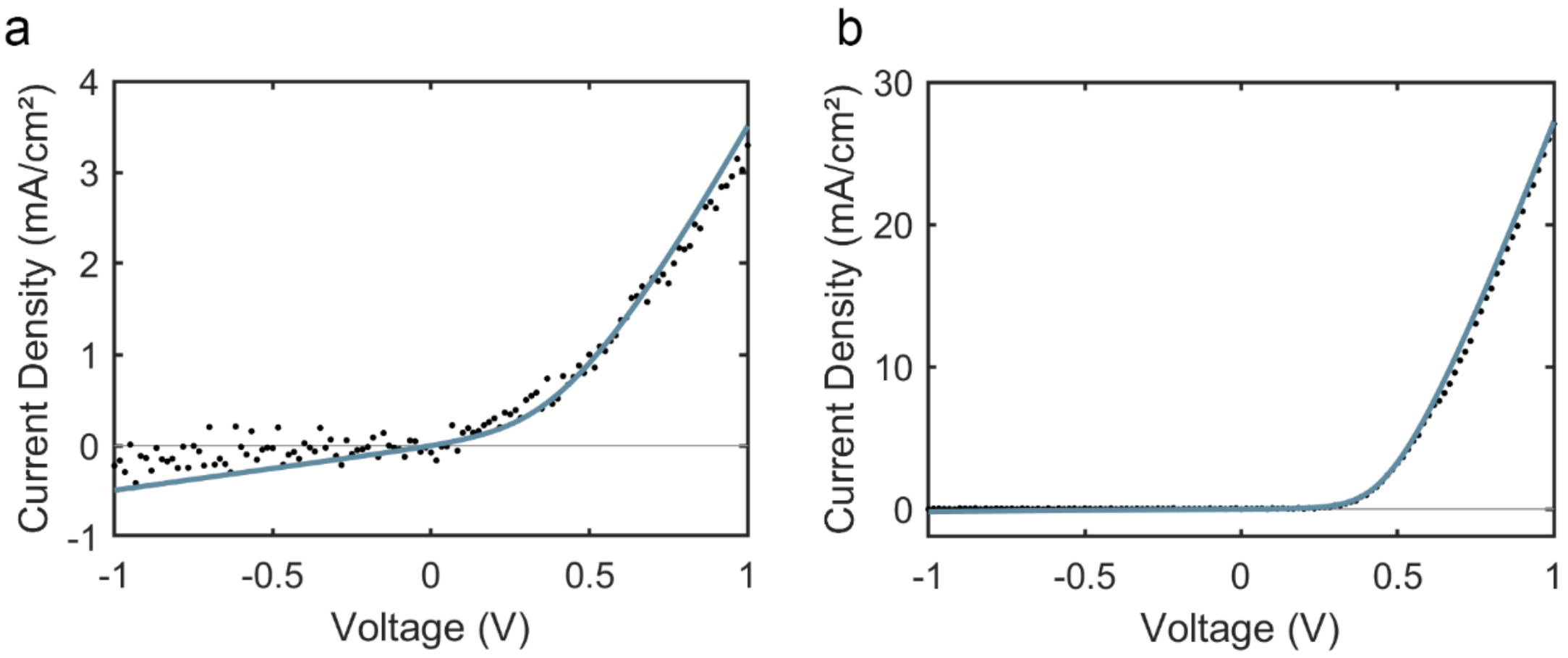

Figure S6: Experimental dark *J-V* characteristics (solid black circles) of the two devices, together with corresponding modelled curves (blue solid lines). The modelled curves for each device are obtained with the same fitting parameters as the illumination curves of the same device, except for the linear series resistance components, which are illumination dependent. (a) Device on $SiO_2$/Si. (b) Device on glass.

**SI_Section 8. Band gap energy determination.**

To determine the band gap energies in Figure 2f we use the relationship between the absorption coefficient $\alpha$ of a semiconductor and the photon energy $h\nu$: $\alpha = K(h\nu - E_G)^m/h\nu$, with $K$ being a constant and $m$ a number characterising the transition type ($m$=1/2 for a direct transition and $m$=2 for an indirect transition).[6] Rearranging yields $(\alpha h\nu)^{1/m} \propto h\nu - E_G$, implying that $E_G$ can be determined from a linear extrapolation of $(\alpha h\nu)^{1/m}$ against $h\nu$. For photon energies around the band gap, $\alpha$ is proportional to the external quantum efficiency (EQE) of the device, so that $\alpha = -\ln(1 - \mathrm{EQE})/d$ where $d$ is the device thickness.

**SI_Section 9. Device micrographs and AFM profiles.**

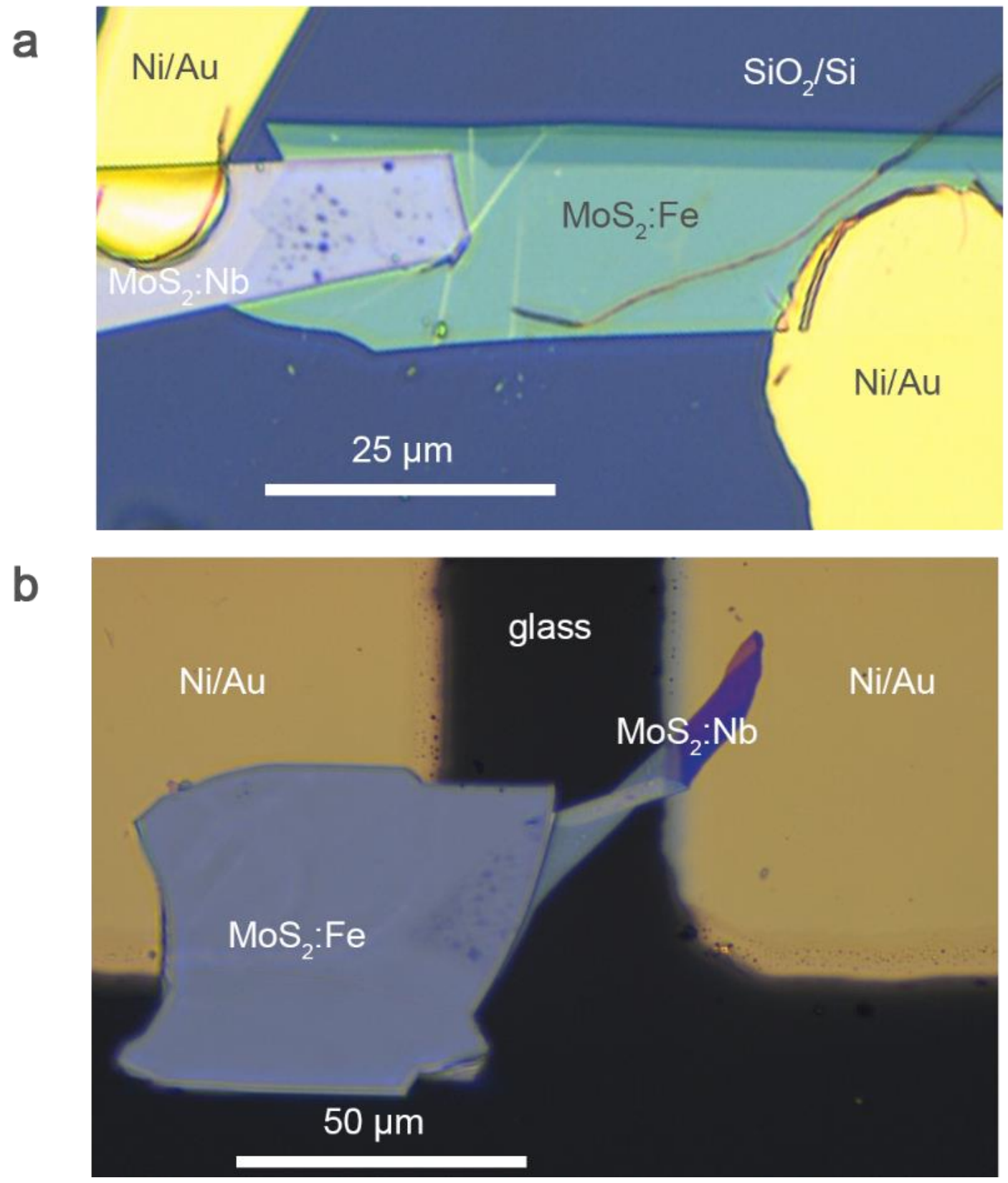

Figure S7: Optical micrograph of the device on $SiO_2$/Si (a) and device on glass (b).

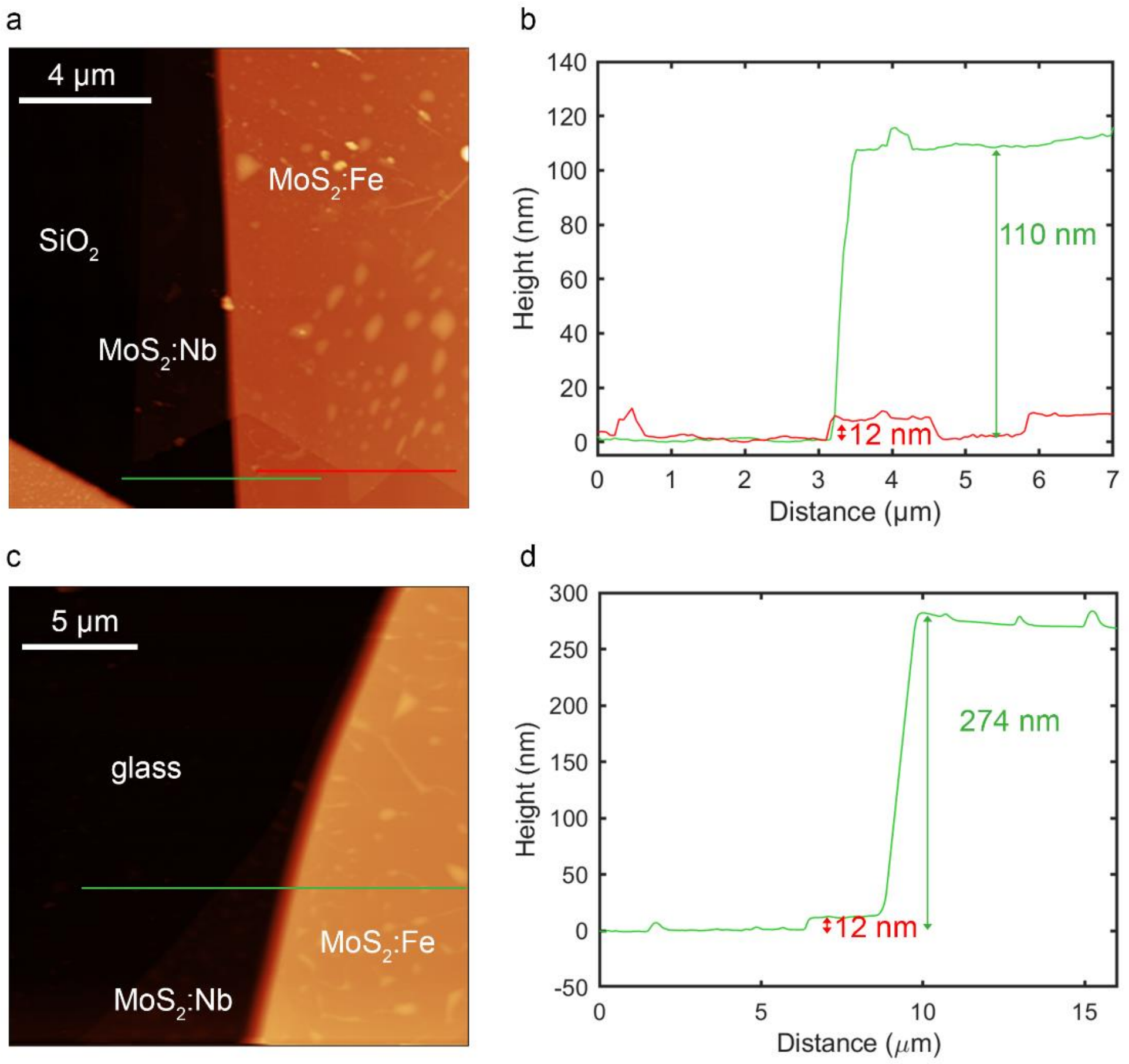

Figure S8: AFM images of the border of the junction area. Device on $SiO_2$ and glass (a,c) with profiles (b,d).

## SI_Section 10: Heterojunction band diagrams

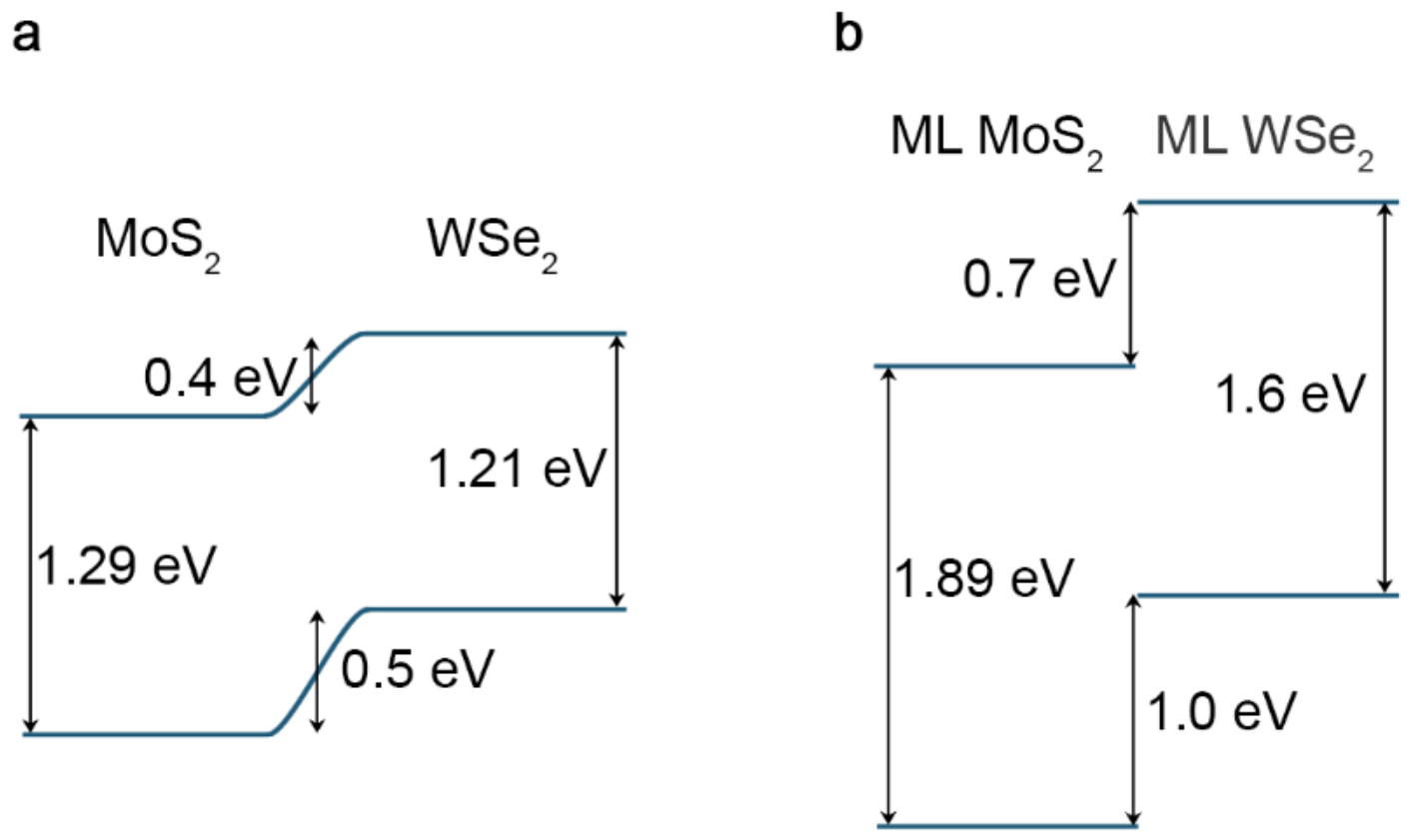

Figure S9: Band diagrams of heterojunctions using intrinsic material of bulk (a) and monolayer (b) $MoS_2$ and $WSe_2$.

## SI_Section 11: Comparison to other works

| Device structure | Highest $V_{OC}$ (V) | Irradiance *P* | Monolayer (ML) / Few-layer (FL) / bulk | Year | Reference |
|---|---|---|---|---|---|
| p-$MoS_2$ / n-$MoS_2$ | 1.02 | halogen, 4 $W/cm^2$ | bulk | 2020 | This work |
| $WSe_2$ / $MoS_2$ | 0.38 | monochromatic 3000 $W/cm^2$ | Few-layer (FL) <15 nm | 2017 | Ref. 3 |
| Gr/$WS_2$/Gr or Gr/$MoS_2$/Gr | 0.18 | monochromatic, unspecified *P* | 5-50 nm | 2013 | Ref. 4 |
| $WSe_2$ / $MoS_2$ | 0.55 | halogen, 0.64 $W/cm^2$ | ML | 2014 | Ref. 14 |
| $WSe_2$ / $MoS_2$ | 0.5 | monochromatic, 100 $W/cm^2$ | ML | 2014 | Ref. 15 |
| $WSe_2$ / $MoS_2$ | 0.27 | monochromatic, unspecified *P* | ML WSe2 and FL $MoS_2$ | 2014 | Ref. 17 |
| $WSe_2$ / $MoS_2$ | 0.32 | monochromatic, 0.037 $W/cm^2$ | FL | 2016 | Ref. 12 |
| $WSe_2$ / $MoS_2$ | 0.5 | monochromatic, 4 $W/cm^2$ | ML | 2018 | Ref. 16 |
| hBN / $MoS_2$ (chemical pn-junction) | 0.5 | monochromatic, unspecified *P* | ML | 2014 | [7] |
| GaTe / $MoS_2$ | 0.224 | monochromatic, 65 $mW/cm^2$ | ML | 2015 | [8] |
| ITO/$MoS_2$/plasma-doped $MoS_2$ | 0.28 | AM1.5G | Bulk | 2014 | [9] |
| $WSe_2$ / $MoS_2$ | 0.36 | halogen, 4.4$mW/cm^2$, | Bulk | 2018 | Ref. 13 |
| $MoSe_2$ / $WSe_2$ | 0.46 | monochromatic, 320 $W/cm^2$ | 3 ML each | 2015 | Ref. 19 |